\documentclass[10pt]{article}

\usepackage{amsmath,amsfonts,bm}

\def\eqref#1{equation~\ref{#1}}

\def\1{\bm{1}}

\DeclareMathAlphabet{\mathsfit}{\encodingdefault}{\sfdefault}{m}{sl}
\SetMathAlphabet{\mathsfit}{bold}{\encodingdefault}{\sfdefault}{bx}{n}

\usepackage[T1]{fontenc}
\usepackage{textcomp}
\usepackage{gensymb}
\usepackage{amsmath} 
\usepackage{color}
\usepackage{graphicx}
\usepackage{xfrac}
\usepackage{soul} 
\usepackage{lineno} 
\usepackage[english]{babel}
\usepackage{blindtext}
\usepackage{amssymb}
\usepackage{caption}
\usepackage{array}
\usepackage{makecell}
\usepackage{comment}
\usepackage{float}
\usepackage{hyperref}
\usepackage{url}
\usepackage{siunitx}

\graphicspath{ {./Figures/} }

\newcommand{\bigO}{\mathcal{O}}
\newcommand{\coup}{\mathcal{C}}

\newcommand{\beginsupplement}{%
        \setcounter{table}{0}
        \renewcommand{\thetable}{S\arabic{table}}%
        \setcounter{figure}{0}
        \renewcommand{\thefigure}{S\arabic{figure}}%
     }

\newenvironment{sciabstract}{%
\begin{quote} \bf}
{\end{quote}}

\title{Ising Model Optimization Problems on a FPGA Accelerated Restricted Boltzmann Machine} 

\author{Saavan Patel,$^{1}$ Lili Chen,$^{1}$  Philip Canoza,$^{1}$ Sayeef Salahuddin$^{1}$}

\author
{Saavan Patel,${}^{1\ast}$ Lili Chen,${}^{1}$ Philip Canoza,${}^{1}$ Sayeef Salahuddin${}^{1}$\\
\\
\normalsize{${}^{1}$Department of Electrical Engineering and Computer Sciences}\\
\normalsize{University of California, Berkeley, California 94720, USA}\\
\\
\normalsize{$^\ast$To whom correspondence should be addressed; E-mail:  saavan@berkeley.edu, sayeef@berkeley.edu}
}

\date{}

\begin{document}

\maketitle













\begin{sciabstract}

In this work we demonstrate usage of the Restricted Boltzmann Machine (RBM) as a stochastic neural network capable of solving NP-Hard Combinatorial Optimization problems efficiently. By  mapping the RBM onto a reconfigurable Field Programmable Gate Array (FPGA), we can effectively hardware accelerate the RBM's stochastic sampling algorithm. We benchmark the RBM against the DWave 2000Q Quantum Adiabatic Computer and the Optical Coherent Ising Machine on two such optimization problems: the MAX-CUT problem and the Sherrington-Kirkpatrick (SK) spin glass. The hardware accelerated RBM shows asymptotic scaling either similar or better than these other accelerators. This leads to $10^7$x and $10^5$x time to solution improvement compared to the DWave 2000Q on the MAX-CUT and SK problems respectively, along with a $150$x and $1000$x improvement compared to the Coherent Ising Machine annealer on those problems. By utilizing commodity hardware running at room temperature, the RBM shows potential for immediate and scalable use. 

\end{sciabstract}

\section*{Introduction}
Combinatorial Optimization problems are a particularly important class of computing problem and are prevalent across disciplines, from scheduling and logistics to analysis of physical systems to efficient routing. These problems belong to the NP-Hard and NP-Complete class, where no polynomial time solution exists. This leads to an interest in novel algorithms, architectures, and systems to solve these problems. The Ising Model problem is an example of this type of problem, with foundations in statistical physics \cite{Barahona1982OnModels,Kirkpatrick1983OptimizationAnnealing,Lucas2014IsingProblems}. For this reason, the Ising Model has emerged as an efficient way of mapping these problems onto various physical accelerators \cite{Yamaoka2016AAnnealing, Inagaki2016AProblems, McMahon2016AConnections, Borders2019IntegerJunctions, Johnson2011QuantumSpins}. Among these, the DWave Adiabatic Quantum Computer \cite{ Blais2000OperationSuperconductors, Johnson2011QuantumSpins} and the Coherent Ising Machine (CIM) \cite{Inagaki2016AProblems, McMahon2016AConnections} are two promising implementations that are capable of solving large scale combinatorial optimization tasks by mapping them onto the Ising model. 

The Boltzmann Machine is a stochastic neural network over binary variables which maps directly onto the Ising model. Because of this, the Boltzmann Machine has received attention for its usage to solve Combinatorial Optimization problems  \cite{Korst1989CombinatorialMachine, Aarts1987BoltzmannApplications}. However, the standard Gibbs Sampling algorithm \cite{Geman1987StochasticImages, Ackley1985AMachines} used with the Boltzmann Machine is computationally expensive due to its sequential nature, and many samples needed to reach convergence.  Restricted Boltzmann Machine (RBM) \cite{Hinton2002TrainingDivergence} can address some of these problems by introducing a parallel sampling scheme via removing intra-layer connections. This is why Restricted Boltzmann Machines have found substantial interest as next generation accelerators for computationally difficult problems \cite{Patel2020LogicallyFactorization, Bojnordi2016MemristiveLearning, Borders2019IntegerJunctions, Sutton2017IntrinsicNanomagnets}. 

In this work we show how the Restricted Boltmann Machine's parallel stochastic sampling scheme is a strong candidate for Ising Model accelerators, especially for combinatorial optimization tasks. We exploit the intrinsic parallelism in this architecture by mapping it onto an FPGA based design with the flexibility in memory and compute to efficiently take advantage of this parallelism. This accelerator allows us achieve better performance than other accelerators based on quantum computation (DWave 2000Q Computer) or novel physical phenomena (Coherent Ising Machine). We demonstrate the performance boost by benchmarking on two types of problems, the Dense Max-Cut Problem up to 200 nodes, and finding ground states for the Sherrington-Kirkpatrick (SK) spin system up to 150 nodes. The CIM and RBM both show a $\bigO(e^{-N})$ performance for probability of reaching the ground state with the DWave annealer demonstrating  $\bigO(e^{-N^2})$ performance. This leads to an asymptotic time to solution performance of the RBM ($\bigO(e^{\sqrt{N}})$) which is better than that of the DWave Annealer ($\bigO(e^{N})$) and similar to that of the CIM (also $\bigO(e^{\sqrt{N}})$) but with a large constant factor scaling improvement over all problem sizes evaluated here.


\section*{Results}

\subsection*{Stochastic Sampling Algorithm}
\label{sec:alg}

The Restricted Boltzmann Machine is a stochastic neural network that encodes an exponential family probability distribution over binary variables, taking the form given in Equation \ref{eq:prob}. The energy function $E(v, h)$ is typically set to reflect the problem being solved, and can take a variety of possible forms \cite{Bremaud1999}. Here we choose the energy function to reflect linear synaptic weights with two body interactions to allow for simplicity of hardware and algorithmic implementation.

\begin{equation} \label{eq:prob}
    p(v, h) = \frac{1}{Z} e^{-E(v, h)}, v \in \{0, 1\}^n, h \in \{0, 1\}^m
\end{equation}

\begin{figure*}[!ht]
\begin{centering}
\includegraphics[width=\linewidth]{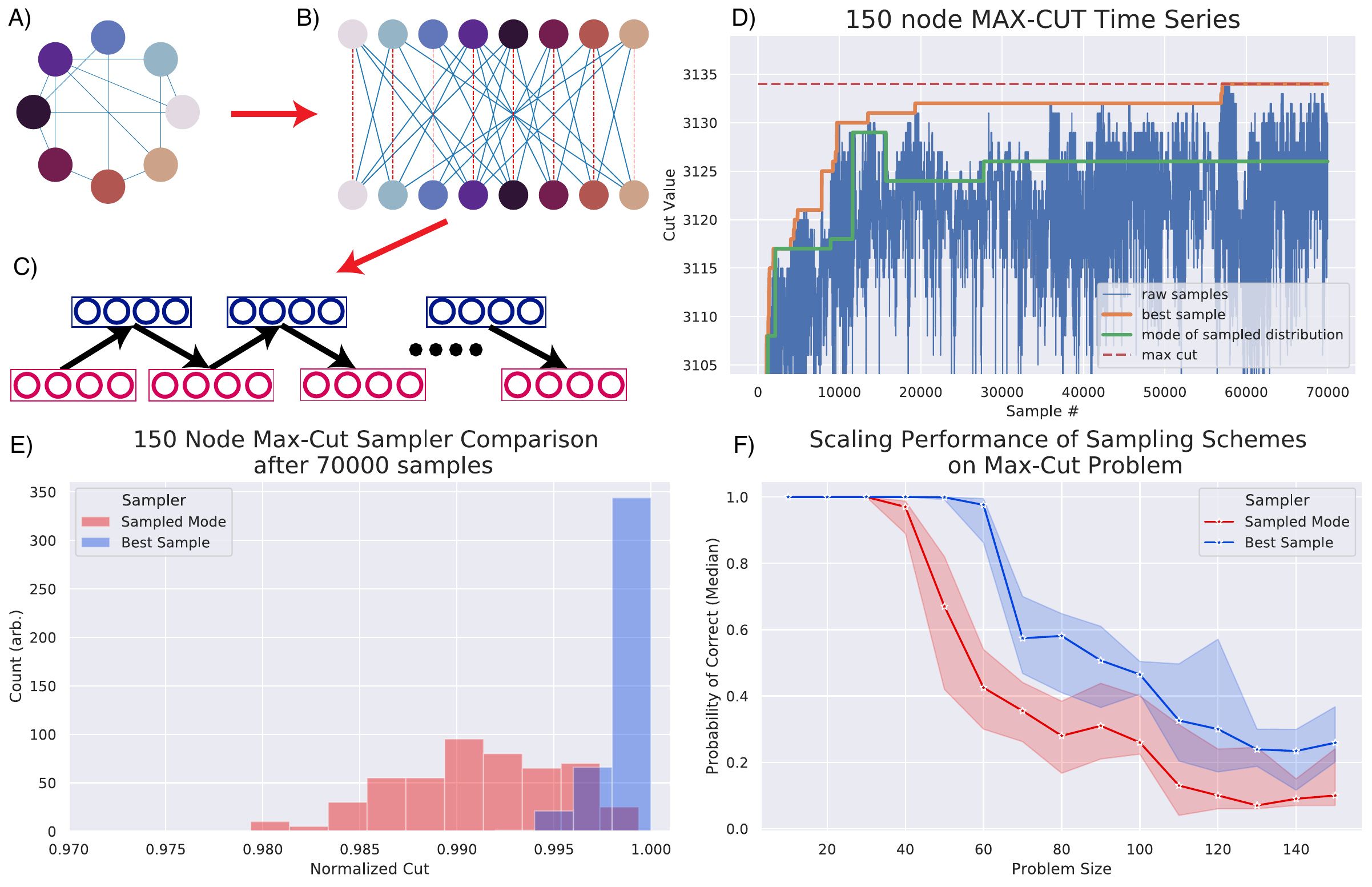}
\par\end{centering}
\caption{\label{fig:alg}. \textbf{Demonstration of RBM structure and sampling algorithm} \\ 
{\bf (A)\/} Structure of the input graph for an Ising Model type algorithm. The graph is fully connected, with no restrictions on the size or magnitude of the weight matrix. 
{\bf (B)\/} The Ising Model is mapped to an RBM by making two copies of each graph node and edge and arranging them into a bipartite graph. One copy is in the ``visible'' layer neurons and one in the ``hidden'' layer neurons, with no intra-layer edges. Each physical copy of the neuron is connected by a ``coupling'' parameter ($\coup$) which constrains the two copies to be the same value. 
{\bf (C)\/} Due to the lack of intra-layer connections, the layers can be sampled in parallel. Each of the neurons in a layer is sampled in parallel and used to calculate the values of the opposite layer, creating a two-step sampling procedure. This sampling procedure proceeds until the output of the algorithm has reached the ground state, or until the algorithm output is of sufficient quality.
{\bf (D)\/ } A demonstrative sampling run showing two different methods for interpreting the output samples from the RBM. 
{\bf (E)\/} A histogram showing the output cuts after 1000 independent sampler iterations with $\coup = 12, N_s = 70000, \beta = 0.25$ on a 150 Node Max-CUT problem, comparing the performance of the two sampling methods. 
{\bf (F)\/} Analysis of the scaling of both of these sampler types. We see that both the sampled mode and best sample procedure perform well with the best sample method performing a constant factor above the sampled mode method.
}

\end{figure*}

As many optimization problems have been formulated for the Ising Model \cite{Lucas2014IsingProblems}, they must be transformed from the fully connected Ising model problem, to the bipartite graph structure in the RBM as demonstrated in Figures \ref{fig:alg} A) and B). To do this, each logical node in the original graph is copied to create two physical nodes in the RBM, with one being in the visible layer and one being in the hidden layer. The connection between the two physical copies is referred to here as the ``coupling coefficient'' ($\coup$) and forces both nodes to remain at the same value. Once the bipartite graph is formed, the RBM energy function is set to $E(v, h) = v^T W h$ with $W$ being the bipartite graph's weight matrix. This encodes the lowest energy states in the original Ising Model problem as the highest probability state in the RBM probability distribution. Further details on the RBM embedding method are outlined in the Methods section. 

To sample from the RBM probability distribution, we perform block Gibbs sampling \cite{Hinton2002TrainingDivergence}, a type of Markov Chain Monte Carlo, on the RBM nodes. Each neuron has a stochastic activation function where $p(v_i = 1 | h) = \sigma(w_i^Th+b_i)$ and $\sigma(x) = (1 + e^{-x})^{-1}$, with $w_i$ being the $i$th row vector of the weight matrix, and $b_i$ being the bias associated with that neuron. The lack of intra-layer connections allows for each neuron in a layer to be sampled in parallel, creating a massively parallel sampling scheme. Each layer is sampled and the result is passed to the next layer to get the next sample. This process of passing neuron activations back and forth is demonstrated in Figure \ref{fig:alg} C). 

An example of the sampling algorithm on a 150 Node MAX-CUT instance is shown in Figure 1 D). The raw samples produced stochastically fluctuate through the high probability states, with a higher probability of transitioning into and staying in the highest probability/lowest energy state. The samples can be analyzed to find the solution in one of two ways; either the samples can be collected and the mode of the sampled distribution is taken as an estimate for the highest probability state, or the unnormalized probability can be calculated for each sample and the highest probability state seen thus far can be taken as an estimate of the highest probability state for the underlying distribution. The first method is related to the ``mixing time'' of the Markov Chain and the second is related to the ``hitting time'' of the Markov Chain \cite{Bremaud1999, Patel2020LogicallyFactorization}. The mixing time method requires less computation but more post-processing, while also allowing for analysis of the full distribution to find other potential high probability states as solutions to the problem. The hitting time method produces only one output solution and converges to the solution faster than the mixing time method, but it requires an extra computation for every sample. In Figure 1E) we show that the hitting time method performs better for the same number of samples, with a cut distribution closer to the maximum. Further, Figure 1F) shows that the hitting time method has a constant factor of improvement in probability of reaching the ground state over the mixing time method. This result is expected from the theory of Markov chain samplers \cite{Oliveira2011MixingChains, Peres2015MixingSets}.

\subsection*{FPGA Acceleration and Hitting Time Engine}
\label{sec:FPGA}
To show the inherent parallelism of the RBM architecture, we map the problem onto an FPGA accelerator. FPGA based Accelerators for the RBM have been made before, but most focus on RBMs in the Machine Learning context \cite{Ly2009AMachines, Kim2009AImplementation, Kim2010AMachines}, or for resource constrained environments \cite{Li2016UsingClassifier}. We base our FPGA accelerator on previous work \cite{Patel2020LogicallyFactorization} which implements a high performance accelerator specifically for fast 
Gibbs sampling operations on an RBM.

The FPGA based accelerator runs at 70$\si{Mhz}$ and produces 1 sample each cycle. Although this runs at a clock frequency considerably slower than a typical CPU or GPU  (usually 1-4$\si{Ghz}$), the FPGA accelerator is able to produce samples at a considerably faster rate than a CPU or GPU implementation \cite{Patel2020LogicallyFactorization}. This speedup can be attributed to the parallelism, binary activations, efficient sigmoid approximations, and reduced precision weights. These cause matrix multiplications to become fixed precision accumulation minimizing the complexity of computation. The accelerator supports 9 bit precision fixed-point weights and biases, which allows for solving Ising Hamiltonians for various problems other than the one included in here. We note that while the problems benchmarked in this work only use weights in $\{-1, 0, 1\}$, we wanted to maintain the generality of our accelerator to solve a variety of other problems within the RBM framework, such as machine learning inference \cite{Hinton2002TrainingDivergence}, and other instances of NP-Hard Combinatorial Optimization problems \cite{Lucas2014IsingProblems, Patel2020LogicallyFactorization}. We note that if we were to restrict the weights our RBM implementation was able to support further we would expect the accelerator as designed to support larger problem instances. 

As outlined above, the two methods for finding the best solution to the optimization problem involves either using the mixing time method or the hitting time method. As the hitting time method has a theoretical advantage for solution quality, we prefer using this. Calculating the probability of each sample on the CPU is computationally expensive, so we implement a ``hitting time engine'' on the FPGA which calculates an approximate, unnormalized log-probability for each sample and keeps track of the state with the highest probability. By using partial computations already available when computing node activations, the hitting time engine has negligible hardware costs and is able to decrease the amount of FPGA to CPU communication to a minimum, freeing the CPU for other computational tasks. The details of the hitting time engine are outlined in the Methods section. 

\subsection*{Effect of Sampler Parameters on Algorithm Performance}
\begin{figure*}

\begin{centering}
\includegraphics[width=\linewidth]{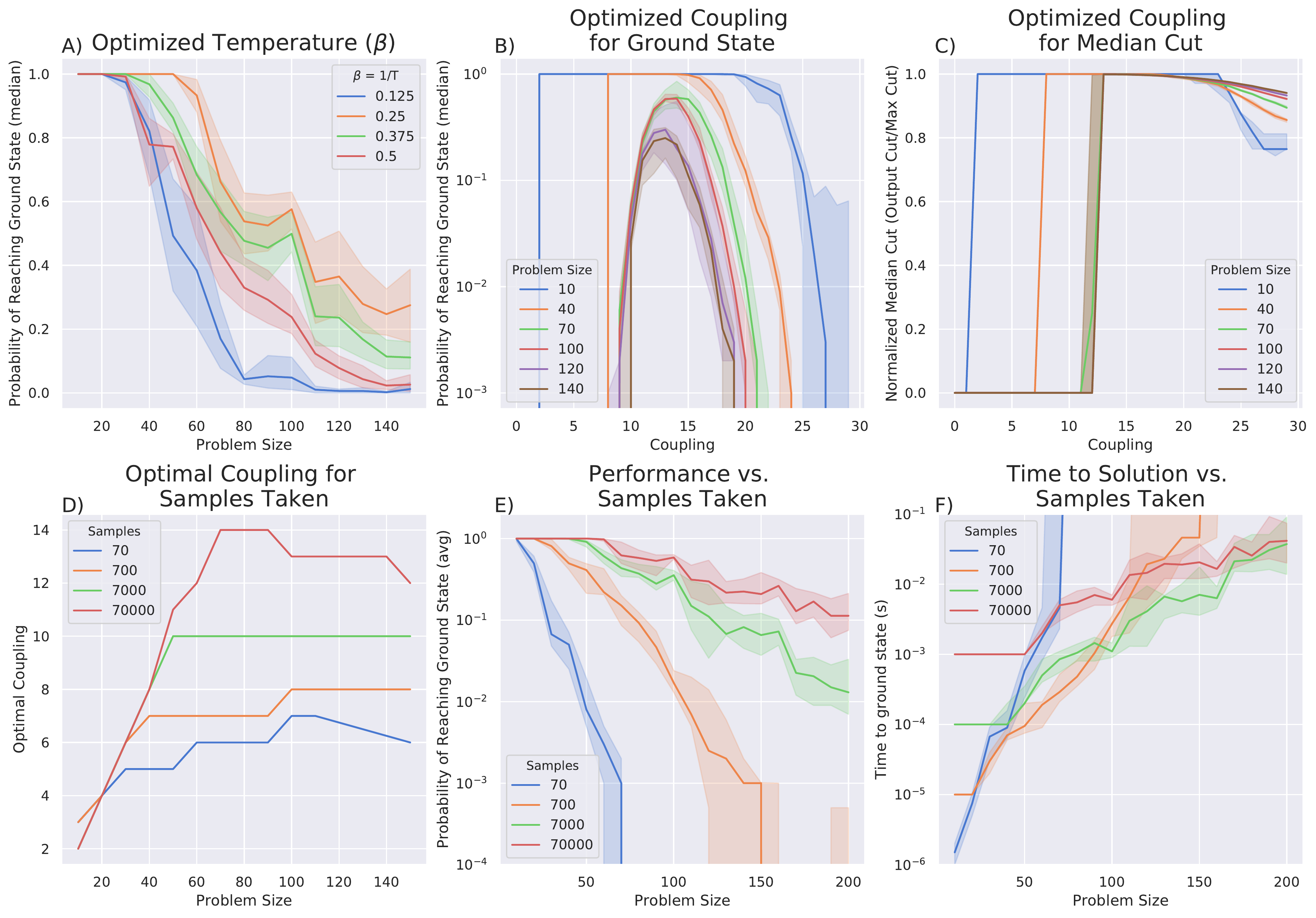}
\par\end{centering}
\caption{\label{fig:par}. \textbf{Optimization of Algorithm Hyperparameters on the MAX-CUT problem} \\ 
{\bf (A)\/} The parameter $\beta$ scales the weights and biases by a constant factor to change the speed of convergence of the sampler. We settle on $\beta = 0.25$ as the optimal parameter, which is used for all experiments in this paper. 
{\bf (B)\/} We show the performance on varying problem sizes for varying coupling parameters at a fixed $\beta=0.25$ and $N_s = 70000$. This shows that for the MAX-CUT problem the $\coup$ is generally optimal at $\coup \approx 12$ for most problem sizes.  
{\bf (C)\/} Although the probability of reaching the ground state is sensitive to the coupling parameter, the median cut outputted from the algorithm tends to be very close to the optimal value for a large range of coupling values. Below a certain value, the median value is very low, but it undergoes a sharp transition to its peak cut value, before slowly degrading. 
{\bf (D)\/} The number of samples taken also increased the optimal coupling value for a given value. This is a consequence of the mixing time of the underlying distribution, where smaller coupling values mix faster but output statistics that are further from the ideal distribution for the problem. 
{\bf (E)\/ } With fewer samples taken, the probability of outputting the ground state decreases significantly. For a given number of samples, the probability of reaching the ground state decreases as $\mathcal{O}(e^{-bN})$ with different coefficients $b$ in the exponent. 
{\bf (F)\/} For a given number of samples taken, we can calculate the time to solution by evaluating Equation \ref{eq:Tsoln} along with the data from (E). The floor of this graph for each problem size is the optimal time to solution for the MAX-CUT problem using the hardware accelerated RBM. 
}

\end{figure*}

The choice of parameters has a strong effect on algorithm performance and should be characterized to select optimal parameters for new problem instances. These parameters are found for each problem and set empirically based on trends seen in the data. The parameters for this sampler are the coupling ($\coup$), the temperature ($\beta$), and the number of samples taken ($N_s$). We first optimize over the temperature parameter, as the other parameters have a strong dependence on the temperature chosen. The temperature parameter ($\beta$) refers to a scaling constant in the probability model, where $p(v, h) = \frac{1}{Z} e^{-\beta E(v, h)}$, and is equal to the inverse temperature seen in the physical Boltzmann distribution. As $\beta \rightarrow \infty$ the distribution becomes more sharply peaked with a larger probability difference between the ground state and first excited state. Conversely, as $\beta \rightarrow 0$, the distribution closer to uniform making it easier to sample from and converge to a solution. The results of the temperature analysis are shown in Figure \ref{fig:par} A), where $\beta \approx 0.25$ yields the best results. We note that we only tested in increments of 0.125, as the fixed precision available on our FPGA accelerator only allowed for increments of $2^{-2}$. More precision is possible, but deemed unnecessary in comparison to the hardware costs. 

The coupling coefficient, $\coup$ is a soft constraint that forces the two copies of the logical node to be the same on the physical RBM implementation. For small values of this constraint, the two copies of the logical node are free from constraint, and an incorrect state will generally be chosen as the ground state. In the case of MAX-CUT, this generally leads to a state of 0 cut, where all nodes are the same value. In the case of the SK Problem, the ground state becomes a random state depending on the problem instance values. The coupling coefficient has a much stronger effect on the performance in the MAX-CUT problem than the SK problem, demonstrated by examining the performance for various values of $\coup$. In Figure \ref{fig:par} B) we see the probability of finding the ground state solution strongly peaks around the optimal $\coup \approx 12$ for $N_s = 70000, \beta=0.25$ for the MAX-CUT problem. For the SK problem, we see less of a sharp peak, with a more gentle decline in performance (shown in Supplementary Figure \ref{fig:Param_SK} A)). In addition, the large values of $\coup$ in the MAX-CUT problem cause a slower mixing time leading to worse performance on the MAX-CUT problem compared to the SK problem \cite{Fischer2014TrainingMachines}. If we instead use the median cut outputted by the sampler as our metric, we see a slightly different picture where the performance is very poor below a certain threshold, peaks at the optimal $\coup$, and then slowly degrades with higher value, which is shown in Figure \ref{fig:par} C). 

Additionally, the optimal $\coup$ tends to change with the number of samples taken, demonstrated in Figure \ref{fig:par} D). This is because small values of the $\coup$ allow for the system to approach the model distribution faster via a smaller mixing time and hitting time \cite{Hinton2002TrainingDivergence, Bremaud1999} but this comes at the cost of having the highest probability state be a state of zero cut (See Supplementary Figure \ref{fig:CoupVsZero} and accompanying discussion for further details). The result is that for small problem sizes, where enough samples are taken such that the sampled distribution is very close to the model distribution, we see a linear relation between the problem size and the optimal $\coup$. This also corresponds to the region where the probability of reaching the ground state, $p_{gnd} \approx 1$ as shown by comparing Figure \ref{fig:par} D) and E). When problem sizes go past this point and the problem's sampled distribution is sufficiently far from the model distribution so it does not correctly identify the mode in all cases, the optimal coupling parameter is mostly flat. 

As the number of samples taken increases, the sampled distribution approaches the model distribution, and the ground state can be correctly identified. This generally causes a smooth and monotonic increase in probability of reaching the ground state as shown in Figure \ref{fig:par} E).  The stochastic hill climbing of the Gibbs sampling algorithm causes the mode to be correctly identified and found well before the full distribution is mapped. Additionally, as each trial is independent, many trials can be performed and a higher success probability can be reached by probability amplification \cite{Motwani1995RandomizedAlgorithms}. We combine these properties with the Time to Solution framework used in other works \cite{Hamerly2019ExperimentalAnnealer, McMahon2016AConnections, Cai2020Power-efficientNetworks, King2018EmulatingAlgorithm}, and adapted for the RBM in equation \ref{eq:Tsoln}, as the standard for evaluating probabilistic accelerators. This corresponds to the 99\% quantile for reaching the ground state of a given problem. 
\begin{equation} \label{eq:Tsoln}
    T_{soln} = \frac{N_s}{f_{clk}}\frac{log(0.01)}{log(1 - p_{gnd})}
\end{equation}

In this equation $N_s$ is the number of samples taken, $f_{clk}$ the clock frequency ($70 \si{Mhz}$ for our FPGA implementation), and $p_{gnd}$ the probability of reaching the ground state for that particular problem. We use this equation along with the data from Figure \ref{fig:par} E) to create Figure \ref{fig:par} F) which shows the time to solution for various $N_s$. We note that if multiple FPGAs are available we can parallelize this scheme and further reduce the $T_{soln}$. Here we use a sequential Time to Solution framework for fair algorithmic comparison to other accelerators. From the graph in Figure \ref{fig:par} E) we see that the optimal number of samples taken (in this framework) is generally lower than the number of samples needed to reach very high accuracy on an individual problem. We can take the lower bound on the data in Figure \ref{fig:par} F) and create a Pareto-optimal boundary for performance on these problems. The same parameter optimizations presented in Figure \ref{fig:par} for MAX-CUT are performed for the SK problem and shown in Supplementary Figure \ref{fig:Param_SK}. 

\subsection*{Benchmarking Performance}

Many accelerators have been developed for solving these Ising Model problems, such as specialized ASICs \cite{Yamaoka2016AAnnealing, Aramon2019Physics-inspiredAnnealer, Boyd2018SiliconNews, Schneider1993AnalogCircuits}, FPGA designs \cite{Belletti2009Janus:Computing, Ko2019Flexgibbs:Graphs}, Memristor based accelerators \cite{Bojnordi2016MemristiveLearning, Wan2020AModels, Cai2020Power-efficientNetworks}, Quantum Mechanical Accelerators based on quantum adiabatic processes \cite{Blais2000OperationSuperconductors, Johnson2011QuantumSpins}, Optical Parametric Oscillators \cite{Wang2013CoherentOscillators, McMahon2016AConnections}, Magnetic Tunnel Junction \cite{Camsari2017StochasticLogic, Camsari2017ImplementingMTJ, Borders2019IntegerJunctions} and many others. In this work we benchmark against two notable candidates: the DWave 2000Q Quantum Adiabatic Computer and the Optical Coherent Ising Machine as they represent the state of the art in two different domains of accelerators. The DWave Annealer has 2000 spins, but due to limited connectivity is only able to solve instances of the dense max-cut and SK problem up-to 62 nodes \cite{ Hamerly2019ExperimentalAnnealer}. The Stanford CIM is able to solve fully connected problems up to 100 nodes \cite{McMahon2016AConnections}, and the NTT CIM is able to solve fully connected problems to 2000 nodes \cite{Inagaki2016AProblems}. The single FPGA instance presented in this work is able to solve up to 200 node optimization problems. 

We benchmark this sampling algorithm on the Dense MAX-CUT problem and the Sherrington-Kirkpatrick (SK) Problem. The MAX-CUT problem involves separating the nodes into two groups and finding the maximal graph cut which separates the two groups of nodes. The MAX-CUT problem is mapped onto the Ising Model by setting weight matrix values $w_{ij} = +1$ with probability $p = 0.5$ and $0$ otherwise. The SK problem sets the Ising Model weights to $w_{ij} = +1$ with $p=0.5$ and $w_{ij}=-1$ otherwise. The problem is then transformed from the Ising Model and fully connected Boltzmann Machine into the RBM using the above methodology to form a probability distribution on the RBM, where the maximum cut for the MAX-CUT problem, or minimal energy for the SK problem, is encoded as the highest probability state. We benchmark on instances from 10 to 200 nodes in increments of 10 on the MAX-CUT problem and 10 to 150 nodes in increments of 10 on the SK problem. Each node value in both problem instances has 10 randomly generated problems, each of which is run 10000 times to generate probabilities of reaching the ground state. Problem instances for node values $\leq 150$ were provided by Ref \cite{Hamerly2019ExperimentalAnnealer}, with instances >150 generated by us. Further details of problem instances are detailed in the Method section. 

\section*{Discussion}
\label{sec:Disc}
\subsection*{Performance Comparison}
\begin{figure*}

\begin{centering}
\includegraphics[width=\linewidth]{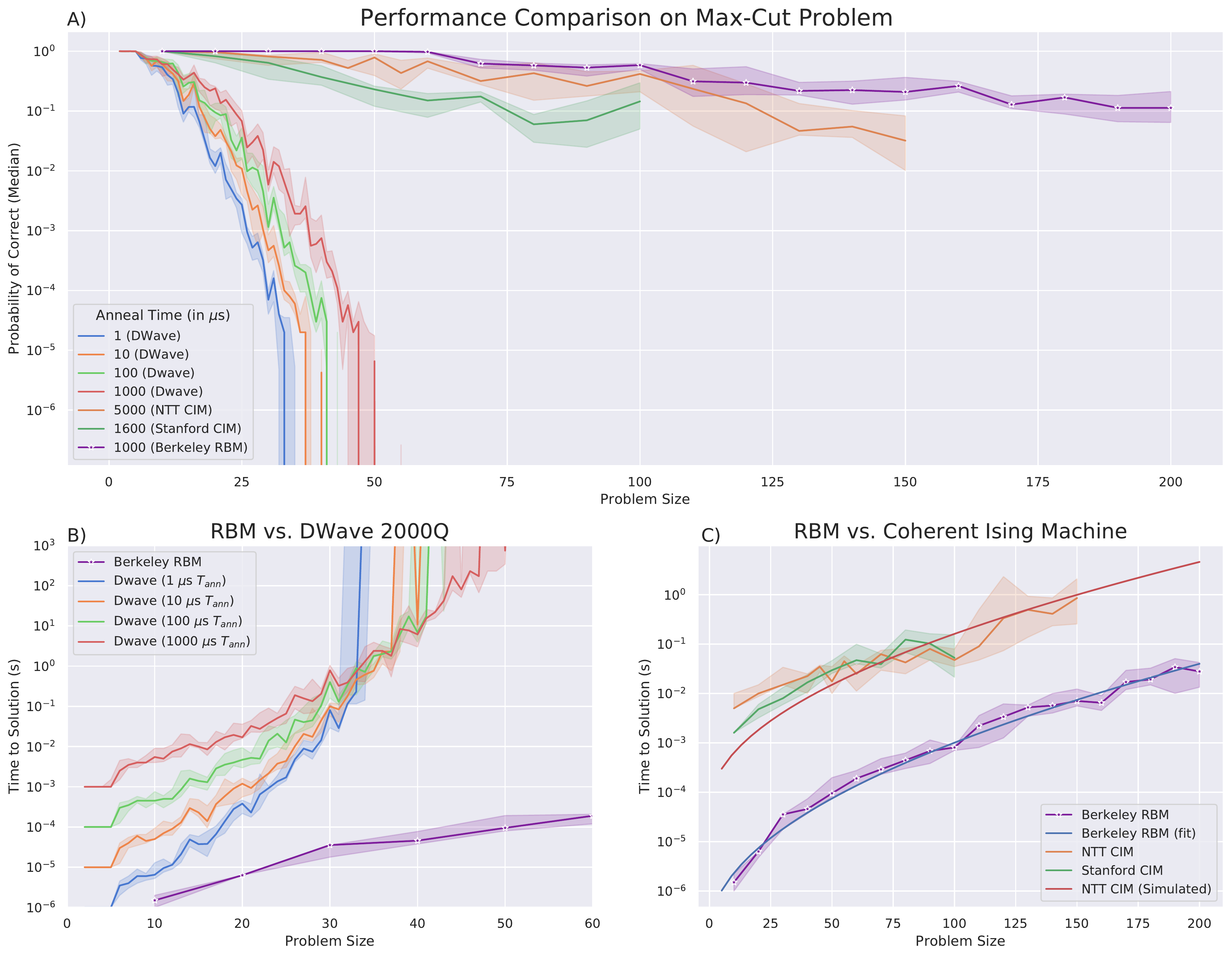}
\par\end{centering}
\caption{\label{fig:MaxCut}. \textbf{Benchmarking and Comparisons on the Dense MAX-CUT Problem} \\ 
{\bf (A)\/} A comparison of performance using the probability of reaching the ground state in various physical annealers as compared to the FPGA accelerated RBM. We see that for similar annealing times, the RBM achieves a best in class probability of reaching the ground state while maintaining a faster annealing time.
{\bf (B)\/} Using the time to solution framework described in Figure \ref{fig:alg} F) and above we compare the performance of Dwave 2000Q to the FPGA accelerated RBM. We see a 7 order of magnitude difference in time to solution for the largest problem instances that the DWave can fit. In addition, we show that the RBM has better scaling properties, with performance differences increasing dramatically with problem size. 
{\bf (C)\/} Comparing the RBM to the Coherent Ising Machine created by NTT \cite{Inagaki2016AProblems} and Stanford \cite{McMahon2016AConnections, Hamerly2019ExperimentalAnnealer} we see a constant factor performance improvement of $\approx 150$x across all problem instances. The RBM shows similar asymptotic scaling to the CIM with both algorithms scaling as $\mathcal{O}(e^{\sqrt{N}})$. }

\end{figure*}

\begin{figure*}

\begin{centering}
\includegraphics[width=\linewidth]{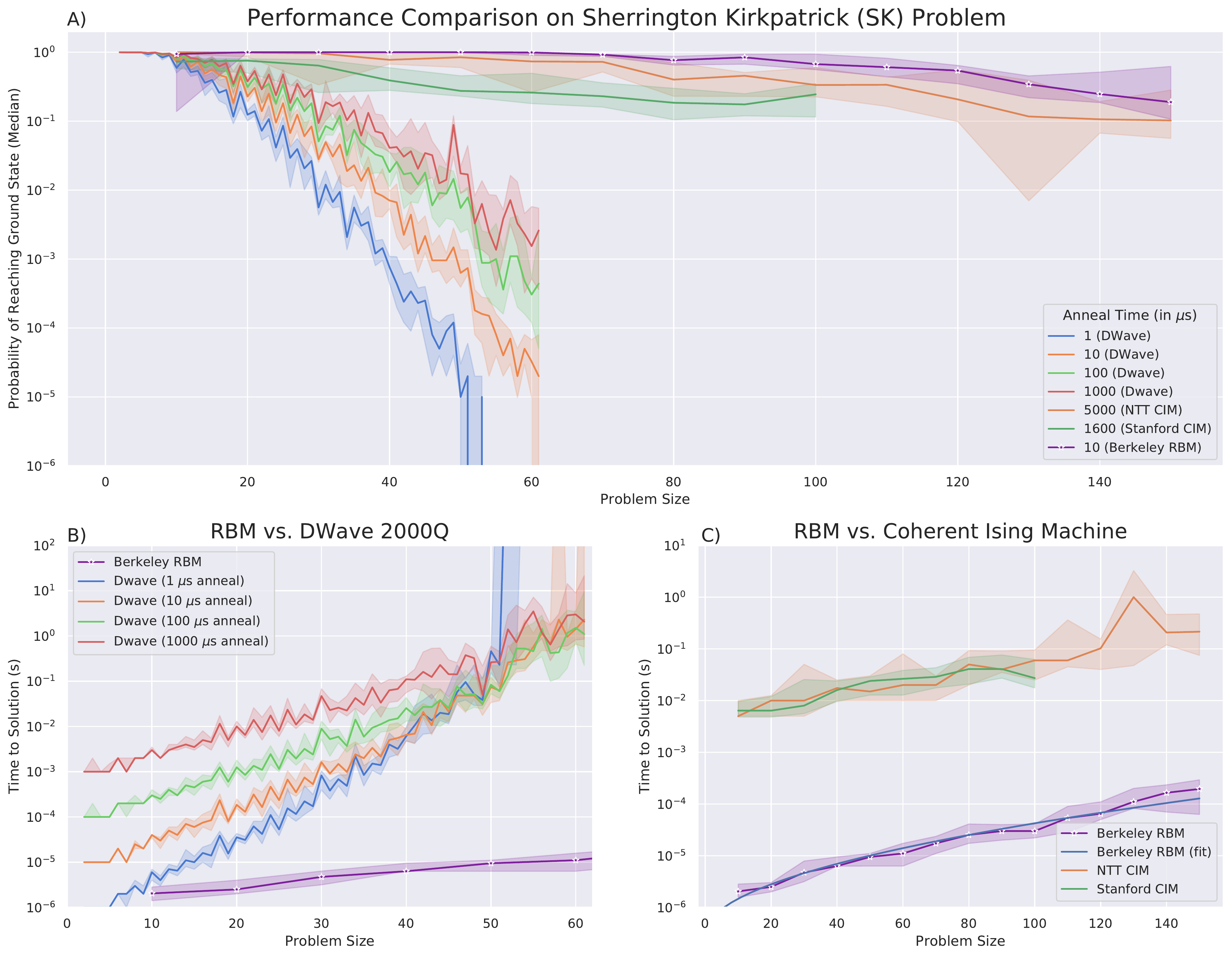}
\par\end{centering}
\caption{\label{fig:SK}. \textbf{Benchmarking and Comparisons on the Sherrington-Kirkpatrick (SK) Problem} \\ 
{\bf (A)\/} Similar to Figure \ref{fig:MaxCut} A), we compare the performance for a fixed Anneal Time on the Sherrington Kirkpatrick. Compared to the MAX-CUT problem, the RBM performs considerably better on this problem instance, only requiring 10 $\mu s$ to get to the ground state with very high probability. This is compared to DWave and the CIM requiring 100x the anneal time to get close to this performance. 
{\bf (B)\/} Compared to the DWave 2000Q, we see a performance increase of $10^5$ on large problem instances with better asymptotic performance on the RBM in these problem instances. The lack of connectivity for the DWave annealer contributes to the drop in performance on these instances as many logical copies need to be made to accommodate the fully connected SK graph. 
{\bf (C)\/} The RBM also compares very favorable to the two instance of the Coherent Ising Machine \cite{Inagaki2016AProblems, McMahon2016AConnections}, with a $1000$x time to solution difference on the largest problem instances. The scaling performance of these two problems also suggests that the RBM will continue its constant factor performance increase for much larger instances of the SK problem. 
}

\end{figure*}

\begin{figure*}[ht]

\begin{centering}
\includegraphics[width=\linewidth]{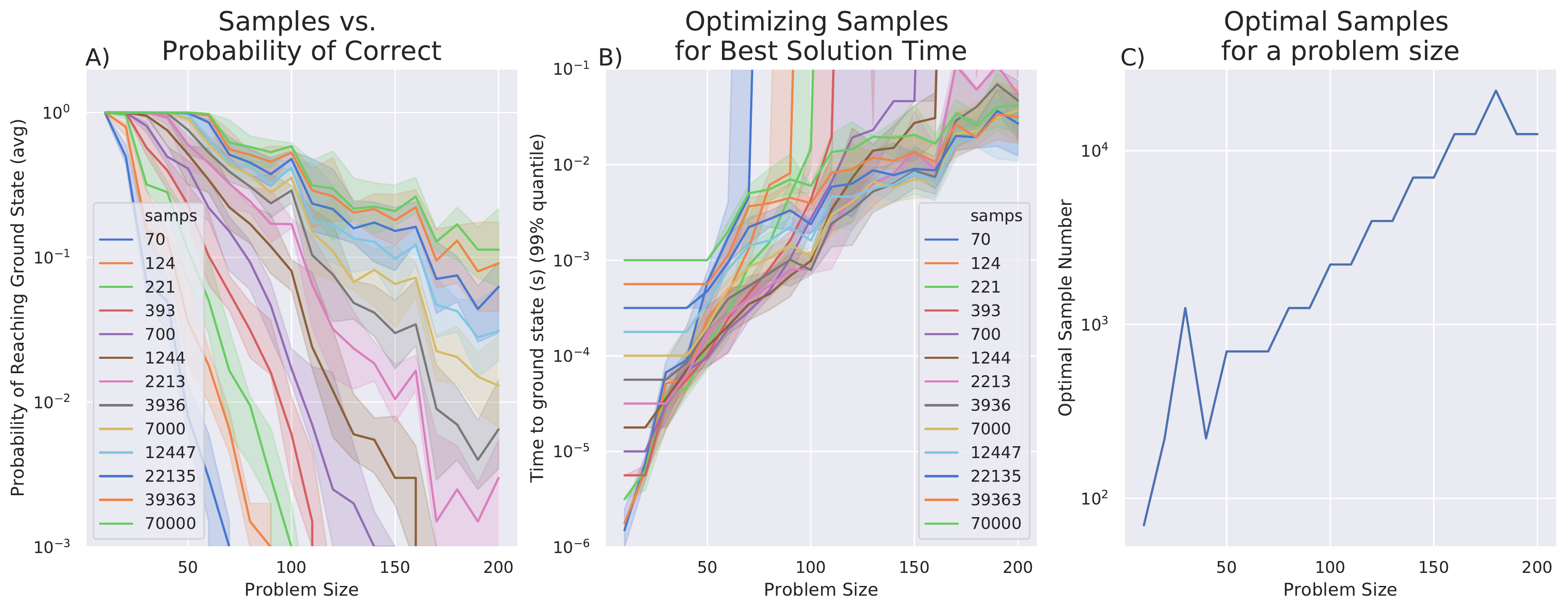}
\par\end{centering}
\caption{\label{fig:SampOpt_MaxCut} \textbf{(Supplementary) Full Analysis of MAX-CUT Sample Optimization} \\ 
{\bf (A)\/} Here we add many more sample values, showing the smooth increase in probability of reaching the ground state as the amount of samples are increased. In this graph, it is also clear that the probability of reaching the ground state decreases as $\mathcal{O}(e^{-N})$, and is approximately exponentially linear. 
{\bf (B)\/} By using the Time to Solution Equation \ref{eq:Tsoln} above, we can see how many iterations are needed to reach the ground state for a given number of samples. We can take the lower bound of the graph at each problem size to find the optimal time to solution for the RBM sampling algorithm as a whole. 
{\bf (C)\/} By finding the sample number for the minimum at each problem size in (B) we can find the optimal sample number for each problem size. This graph shows the $e^N$ increase in optimal sample number as problem size increases. The discrete jumps shown here are due to the discrete number of samples taken, and would become smooth as this procedure is interpolated for more sample values. 
}

\end{figure*}

In Figures \ref{fig:MaxCut} and \ref{fig:SK} we show results of benchmarking performance on the MAX-CUT and SK problems. In Figure \ref{fig:MaxCut} A) we show how the probability of reaching the ground state scales for a fixed annealing schedule. We fix $N_s = 70000$ for the RBM, corresponding to $1000 \si{\us}$ of time at $70 \si{Mhz}$ and see that the performance outperforms the other annealers at all problem sizes when given less time to solution. Using the Time to Solution framework shown in Equation \ref{eq:Tsoln} we convert the optimal sampling solution from Figure \ref{fig:par} F) and Supplementary Figure \ref{fig:SampOpt_MaxCut} B) to compare the time to solution against the DWave 2000Q in Figure \ref{fig:MaxCut} B) and the Coherent Ising Machine instances in Figure \ref{fig:MaxCut} C). We see a particularly stark difference in scaling performance when comparing to the DWave 2000Q, where the performance on problem instances drops quickly to 0 after 50 Nodes in the MAX-CUT problem, while the RBM is still able to solve larger instances. When looking at time to solution, this accounts for a $10^6$x difference in performance at 50 nodes. When comparing to the the Coherent Ising Machine, we see very similar scaling performance for time to solution. While the Stanford CIM and the NTT CIM perform very similarly, the RBM performs at a constant $\approx 150$x advantage over all problem sizes. When comparing to the simulated curve, this constant advantage becomes even more apparent. 

This difference in performance for a given problem size is more pronounced when examining the SK problem. First, we note that with much less computation time ($10 \si{us}$ compared to  $\approx 1000 \si{us}$) the RBM is able to outperform both the DWave and CIM for the given problem instances shown in Figure \ref{fig:SK} A). This difference becomes more apparent when looking at the Time to Solution metric, where we see a $10^5$x improvement at 60 nodes compared to DWave in Figure \ref{fig:SK} B) and a constant $10^3$x improvement against the Coherent Ising Machines in Figure \ref{fig:SK} C). As with the MAX-CUT problem, there appears to be a scaling difference between the RBM and the DWave 2000Q, while the difference between the RBM and Coherent Ising Machine seems to be a constant factor improvement.  

\subsection*{Scaling and Connectivity}
One of the biggest challenges in implementing Ising Machines is creating all-to-all connectivity between nodes. When mapping arbitrary graphs onto the Ising Machine, this is a necessary requirement to build a usable machine. The CIM supports this kind of all-to-all connectivity, while the DWave 2000Q, with its limited connectivity, uses $\approx N^2/\kappa$ where $\kappa$ is the level of connectivity in the physical graph \cite{Hamerly2019ExperimentalAnnealer}, leading to a large overhead in computation. A consequence of this is that the DWave annealer has scaling performance of $\bigO(e^{-N^2}))$ in probability of reaching the ground state, while both the CIM and the RBM based methods exhibit $\bigO(e^{-N})$ \cite{Hamerly2019ExperimentalAnnealer, Rnnow2014DefiningSpeedup}. The RBM based methodology, although mapping $N$ logical nodes to $2N$ physical nodes, does not suffer from the same scaling problems as the DWave Machine. We believe this is because the RBM increases the number of physical nodes by a constant factor, rather than a factor that depends on the size of the input graph. 

We expect similar asymptotic performance for both the SK problem and MAX-CUT as they both are NP-Hard problems, based on the Ising Model glassy spin system configuration, using the same embedding and underlying sampling algorithm. This is confirmed based on our experiments with both problems having an underlying $\bigO(e^{-N})$ for probability of reaching the ground state for a fixed number of samples (see Figures \ref{fig:par} E), Supplementary Figure \ref{fig:SampOpt_MaxCut} A), and Supplementary Figure \ref{fig:Param_SK} C)). Based on the same scaling behavior for the ground state probability metric, we see a scaling of $\bigO(e^{\sqrt{N}})$ for both the MAX-CUT and SK problems and fit curves to both problems (see Figures \ref{fig:SK} C) and \ref{fig:MaxCut} C))). We note that as the SK problem requires such few samples for computation, the optimal sample point does not transition very much between different sample values. This causes the behavior to appear linear in the exponential at first glance, but we expect it to follow the $\bigO(e^{\sqrt{N}})$ behavior for larger problem sizes. 

Although we empirically see a probability scaling of $\bigO(e^{-N})$ and a time to solution scaling of $\bigO(e^{\sqrt{N}})$ in the RBM annealing algorithm, we acknowledge that without a theoretical result, this scaling is not proven. More work is necessary to understand the sampling algorithm in greater detail. It should be noted, however, that  the scaling principles seen in the CIM and DWave are also empirical \cite{Hamerly2019ExperimentalAnnealer, Albash2018DemonstrationAnnealing}, thus only experimental data can be reasonably compared and care should be taken when extrapolations are taken of the data.

\subsection*{Effect of $\coup$ and graph embedding on algorithm performance}
The purpose of the coupling constraint ($\coup$) is to enforce the two physical copies of the logical node to be the same value. When the constraint is violated, the two physical copies have different values and the state energy is no longer proportional to the Ising Energy of the problem Hamiltonian being solved. This would imply large values of $\coup$ would improve performance, but that is not what is observed. When the value of $\coup$ is too large, the Markov Chain does not mix quickly and settles in local minima \cite{Hinton2002TrainingDivergence, Fischer2014TrainingMachines, Bremaud1999}. 

As shown by comparing Figures \ref{fig:MaxCut} and \ref{fig:SK}, we can see that the there is a significant difference in performance between the MAX-CUT and SK problems, even when comparing the performance difference to the other annealers. This is partially caused by the difference in optimal $\coup$ required for the problems. While the SK problem uses $ \coup \approx 1$ for all problems, the MAX-CUT problem has an optimal value of $\coup \approx 12$ for the same $N_s = 70000, \beta = 0.25$. In addition, the graph embedding we use leads the MAX-CUT problem to have a high probability for a state with a cut of 0, a state which is suppressed for high values of the coupling parameter (See Supplementary Figure \ref{fig:CoupVsZero} and \ref{fig:ExcludeZero} for further discussion). We would expect that remapping the MAX-CUT problem to the RBM via a different method that requires smaller $\coup$ could result in a increase in performance due to lower mixing times. 

\subsection*{Conclusion}
By exploiting the intrinsic parallelism present in the RBM architecture on a flexible FPGA based accelerator, we show that our sampling framework is competitive with state of the art computing machines based on novel physics. Importantly, we empirically show that there appears to be no scaling advantage for the accelerators based on novel physics, indicating that classical hardware is sufficient for solving the computationally difficult problems chosen here. We additionally show that the RBM has a large constant factor performance advantage on both of these problems.  Although we chose the MAX-CUT and SK problems to benchmark and solve, all of Karp's 21 NP-Complete problems can be mapped onto the Ising Model, and the RBM using the proposed framework, in polynomial resources \cite{Karp1972ReducibilityProblems, Lucas2014IsingProblems, Aramon2019Physics-inspiredAnnealer}. In addition, our usage of a FPGA framework with up to 9 bit precision arithmetic allow for many varieties of these problems to be solved, including arbitrary real world instances that the lower precision DWave and CIM would not support. The use of commodity hardware working at room temperature in a standard server setup allows for widespread adoption and usage. Further accelerator-level parallelization and  scaling also become possible through the use of multi-FPGA designs and communication \cite{Lo2011BuildingMPI, Kim2010AMachines, Ly2009AMachines}, time division multiplexing \cite{Yamamoto2017AFPGAs} and more efficient pipeline stages \cite{Kim2014ANetwork}. 

The work presented represents a proof of concept for the possibility of using parallel, stochastic computing to solve NP-Hard and NP-Complete problems. As these problems represent some of the hardest for traditional computers to solve, this approach has far reaching consequences in fields like logistics, scheduling, resource allocation, and many others. Further improvement on this methodology can be accomplished through the use of novel devices \cite{Camsari2017StochasticLogic, Camsari2017ImplementingMTJ, Borders2019IntegerJunctions, Bojnordi2016MemristiveLearning, Wan2020AModels} for either the matrix multiplication/accumulation or the noise generation. Recently, there has been similar work  which modifies the Hopfield Network with noise and variation from memristors to create stochastic behavior \cite{Cai2020Power-efficientNetworks} creating a network with very similar properties to a Stochastic Boltzmann Machine, showing how the combination of stochastic sampling and a parallel accelerator can yield impressive performance gains. A dedicated accelerator for the RBM using these novel technologies would enable large scale optimization at high speed and throughput, with potential to solve some of the most difficult computational problems. 


\clearpage
\newpage
\section*{Methods}

\subsection*{Mapping Ising Problems onto the RBM}
Ising model problems take the form of a minimization on of an energy function ($E(s)$) over bipolar states ($s \in \{-1, 1\}^n$), that usually takes the form of Equation \ref{eq:Ising}. We will denote quantities relating to the original Ising model with an $I$ subscript, and quantities related to the RBM with an $R$ subscript (i.e. $n_R$ is the number of visible nodes in the RBM, while $n_I$ is the number of nodes in the original Ising model). 

\begin{equation} \label{eq:Ising}
E_I(s) = \frac{1}{2}\sum_{i=1}^{n_I}\sum_{j=1}^{n_I}J_{ij}s_is_j + \sum_{i=1}^{n_I}a_is_i
\end{equation}

The RBM however maps a probability distribution over binary variables ($v \in \{0, 1\}^n$, $h \in \{0, 1\}^m$) without intra-layer connections. This takes the form shown below. 
\begin{align}
    p_R(v, h) & = \frac{1}{Z}e^{-E(v, h)} \\
    E_R(v, h) & = -(\sum_{i = 1}^{n_R}\sum_{j = 1}^{m_R} w_{ij} v_i h_j + \sum_{j = 1}^{m_R}b_jh_j + \sum_{i = 1}^{n_R}c_iv_i) 
\end{align}

The RBM has twice the number of nodes as the original Ising model graph, with $n_{R} = m_{R} = n_{I}  = n$, and each visible node having a corresponding hidden node that should hold the same value in the ground state of the RBM model (i.e. $v_i = h_i$ for the RBM ground state). With this, we can set the weights by equating the energy of an Ising Model state to the the energy in an RBM state (causing minimal energy Ising states to be maximal probability RBM states), and collecting terms related to the same logical node, shown below.  We first show results for an RBM with bipolar states, and then show the transformation between bipolar states and binary states. 
\begin{align}
    E_{I}(s) & = E_{R}(s, s) \\
    \frac{1}{2}J_{ij}s_is_j & = -(w_{ij} s_i s_j + w_{ji} s_j s_i) \\
    a_i s_i & = -(b_i s_i + c_i s_i)
\end{align}
Based on this equality, we set each $w_{ij} = w_{ji} = -J_{ij}$ and $b_i = c_i = -\frac{1}{2}a_i$. This sets the correct probabilities for an RBM for states where $v_i = h_i$, but the probabilities for states where $v_i \neq h_i$ should be correctly penalized. To do this we set $\forall i, w_{ii} = -\coup$, which penalizes states so that $E_R(s, x) \ll E_R(s, s), \forall x \neq s$. The size of $\coup$ is problem specific and is empirically analyzed to find the optimal parameter for a particular problem instance. 

The above transformation works for RBMs with bipolar states ($\{-1, 1\}$), but RBMs traditionally use an energy function over binary variables ($\{0, 1\}$). The transformation between these two is straight forward, and shown below for an arbitrary Ising model. In the below equation $W^*, b^*$ correspond to Ising models where $s\in \{0, 1\}^n$,  $W, b$ are the original weights and biases for the Ising model with $s\in \{-1, 1\}^n$, and $\vec{1}$ is a vector of $1$s. 
\begin{equation} \label{eq:IsingTransform}
    W^* = 4W, b^* = 2(b - W\vec{1})
\end{equation}

Depending on the type of accelerator architecture available, mapping between bipolar and binary states is trivial to perform, and represents minimal computational overhead. In this work, the RBM accelerator was designed to use binary states, and the Ising model problems were adapted for that.

\subsection*{Hitting Time Engine}
The hitting time engine calculates the approximate log probability for each sample that the stochastic sampling algorithm outputs and keeps track of the highest probability state seen so far. This method works to offload the computation of calculating probabilities or aggregating results from the CPU to the FPGA. The FPGA can operate in raw sample output or hitting time engine output, where it either pushes the raw samples to the CPU or the highest probability state seen. The hitting time engine uses partial computations from each cycle to reduce computational overhead for the FPGA. To do this, we first look at the log probability for a given visible node state. 

\begin{align*}
    p(v, h) & = \frac{1}{Z}e^{-E(v, h)} \\
    & = \frac{1}{Z}e^{\sum_{i = 1}^{n}\sum_{j = 1}^{m} w_{ij} v_i h_j + \sum_{j = 1}^{m}a_jh_j + \sum_{i = 1}^{n}b_iv_i} \\
    p(v) & = \sum_h p(v, h) \\
        & = \frac{1}{Z} \prod_{i = 1}^{n} e^{b_iv_i} \prod_{j = 1}^{m} (1 + e^{\sum_{i = 1}^{n} a_j +  w_{ij} v_i} ) \\
log(p(v)) & = \sum_{i = 1}^{n} b_iv_i +  \sum_{j = 1}^{m} log(1 + e^{\sum_{i = 1}^{n} a_j +  w_{ij} v_i} ) - log(Z)
\end{align*}

The $log(Z)$ term is a normalizing constant and can be ignored if we are only comparing probabilities between samples. Additionally, the $log(1 + e^x)$ term is simplified as follows. 
\begin{equation*}
    log(1 + e^{x} ) = 
        \begin{cases}
          x, & \text{if}\ x \ge 0 \\
          0, & \text{if}\ x < 0
        \end{cases}
\end{equation*}

This simplification is valid for $x \gg 0$ and $x \ll 0$, but introduces errors when $x \approx 0$. These errors are not significant in the probability calculation, as the largest contributions to the probability mass are for $x \gg 0$. The $\sum_{i = 1}^{n} a_j +  w_{ij} v_i$ is calculated each cycle to update the hidden units and is thus recycled for calculation of the overall log probability of the given state. This means the only calculation the hitting time engine has left to do is accumulation of the visible biases and accumulation of the thresholded sums that have been pre-calculated by the hidden neurons. The hardware overhead for the hitting time engine is very small due to the efficient usage of these pre-calculated sums. The hardware usage translates to $<1\%$ of additional FPGA LUT utilization and $<1\%$ of additional FPGA flip flop usage (see Supplementary Table \ref{tab:utilization}). 

\subsection*{FPGA Programming}
All programming was done using the Xilinx Vivado suite on the on the Xilinx Virtex UltraScale+ XCVU9P-L2FLGA2104 following Ref \cite{Patel2020LogicallyFactorization}. All weights and biases were stored in on-chip SRAM and communication done over PCIe. The design from that work was slightly modified to widen the bit count from 8 bits to 9 bits. This was used to test effects of various parameters and have sufficient dynamic range to perform experiments. Along with this, partial sums were stored for use in the hitting time engine. 

The hitting time engine is split into two modules, each calculates the log probability for every other cycle. Each module is composed of an accumulator which takes the partial sums from the hidden nodes and accumulates half of them each cycle along with the visible biases. By splitting the calculation over two clock cycles we are able to meet the $70 \si{Mhz}$ timing requirements set by the rest of the design. The hardware cost of these accumulators is approximately the same as an additional visible node (see Supplementary Table \ref{tab:utilization}). We additionally add an extra function in the hitting engine, where we ignore any output that corresponds to a zero cut state. The results of this additional feature are shown in Supplementary Figure \ref{fig:ExcludeZero}, and function to increase the average output cut from the sampler. 

\subsection*{Problem Instances and Validation}
Problem instances for MAX-CUT and SK and performance data for the CIM and DWave accelerators for sizes $\le 150$ were provided by Ref \cite{Hamerly2019ExperimentalAnnealer} and validated by them. Their ground state results are in agreement with ours, as in many runs we were unable to find states with Ising Energies lower than the ground state solutions provided. For the problem instances we generated on MAX-CUT instances, we generated random graphs with edge density $0.5$ and tested the MAX-CUT instances on our solver. As exact solvers, such as BiqMac \cite{Wiegele2007BiqSize, Rendl2007ARelaxations}, time out for instances of size $> 100$ nodes (corresponding to 3 hours of computation time) we were unable to confirm a provable ground state solution to these instances. Instead, we compared the best solution generated by the RBM based solver with best output from the solver implemented in Ref \cite{Isakov2014OptimizedGlasses} and found agreement between those two solutions. As the best solution instances from these two solvers are in agreement and follow the curve that from smaller problem instances, we are fairly certain the RBM is finding the ground state solution. 

To characterize performance of the RBM on the MAX-CUT and SK, each problem instance was run 10000 times and the algorithm output was checked against the ground state solution for that problem. Each problem size had 10 randomly generated instances to test on and results were aggregated based on their performance. Error bars on each of the graphs were generated by calculating the bootstrapped, two tailed, 95\% confidence interval for the parameter being estimated. Solution time calculations for the RBM do not include pre-processing or post-processing of sample data, only the raw computation time required to solve the underlying problem. This is consistent with the methods used for the other annealers. Anneal times for the CIM use the base anneal time without a parallel sampling protocol to fairly compare to the RBM instances and allow for analysis of scaling.



\newpage
\bibliography{References/references.bib}
\bibliographystyle{ScienceAdvances}



\newpage

\noindent \textbf{Acknowledgements:} 
%
This work was supported by ASCENT, one of six centers in JUMP, a Semiconductor Research Corporation (SRC) program sponsored by DARPA. We thank Ryan Hamerly for providing the raw data and problems for comparison to the Coherent Ising Machines, DWave computer and Parallel Tempering solver. \\
\noindent \textbf{Author Contributions} Model Synthesis and Analysis was performed by S.P. and L.C.; FPGA Programming was performed by S.P and P.C.; Manuscript was co-wrote by S.P, L.C, P.C and S.S; S.S supervised the research. All authors contributed to discussions and commented on the manuscript. \\
\noindent \textbf{Competing Interests} The authors declare that they have no competing financial interests.\\
\noindent \textbf{Data and materials availability:} Additional data and materials are available on reasonable request from the authors.  \\
\noindent \textbf{Correspondence} Correspondence and requests for materials can be addressed to either S.P. \\ (saavan@berkeley.edu) or S.S. (sayeef@berkeley.edu).

\onecolumn
\clearpage
\section*{Figures and Tables}

\clearpage

\clearpage

\clearpage

\onecolumn
\clearpage
\section*{Supplementary Material}
\beginsupplement

\clearpage

\begin{figure*}

\begin{centering}
\includegraphics[width=\linewidth]{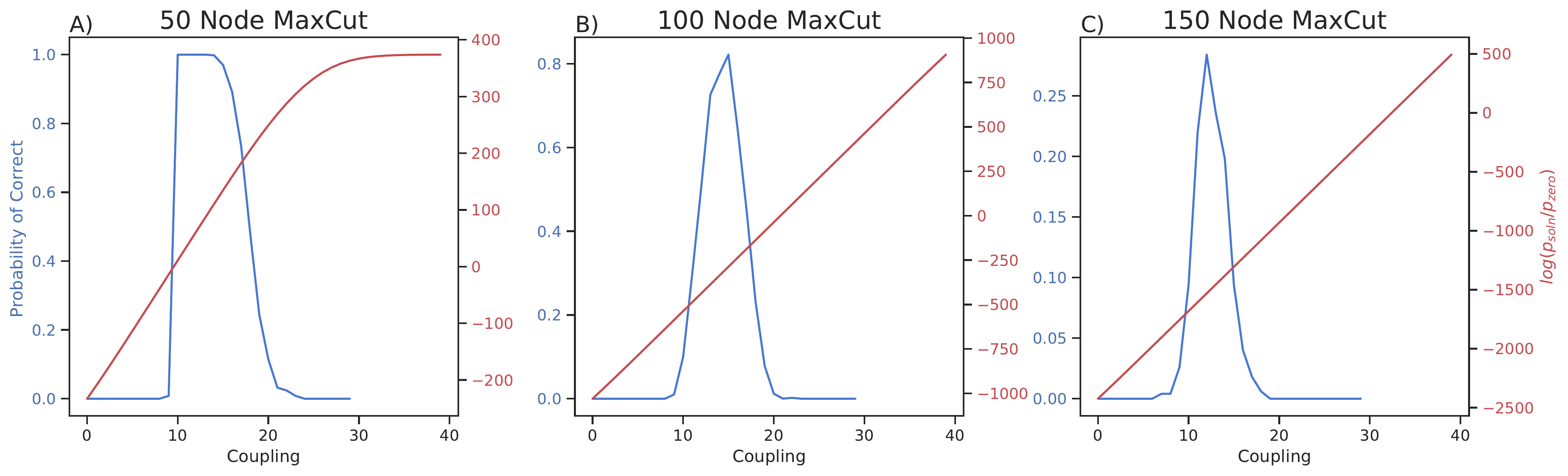}
\par\end{centering}
\caption{\label{fig:CoupVsZero}. \textbf{(Supplementary) Effect of Coupling Coefficient on Sampled Performance vs. Limiting Performance} \\ 
This shows how the relative probability of the zero state (right axis) compares to the probability of reaching the ground state after 10000 iterations (left axis) for $N_s = 70000, \beta = 0.25$ on various problem sizes.
{\bf (A)\/} For $N=50$, we can tell that the $N_s$ is large enough such that the sampler has fully mixed, as the sampler performance peaks when $p_{soln} > p_{zero}$ and the probability of correct $\approx 1$. 
{\bf (B)\/} For $N=100$, we see that the sampler is further from convergence as the peak performance no longer approaches 1, and the optimal coupling parameter is for a state where $p_{soln} < p_{zero}$. 
{\bf (C)\/} As the problem size increases to $N=150$ we see that the sampler is even further from convergence as $p_{soln} \ll p_{zero}$ at the optimal coupling value. 
}

\vspace{2em}
For the MAX-CUT problem there is an intuitive explanation for the role of the coupling parameter. When $\coup = 0$, the two physical copies of a node in the original graph are completely disconnected resulting in the two states not having any direct effect on each other. The maximum cut in this degenerate graph is the one which separates the hidden from the visible nodes and passes through all of the edges and where $v = \{0\}^N, h = \{1\}^N$ or $v = \{1\}^N, h = \{0\}^N$. We refer to this state as the ``zero cut state'' as it corresponds to a state that has zero cut in the original Ising Model graph. As $\coup$ increases, the relative probability of this state decreases compared to the actual MAX-CUT state for the original Ising Model graph. However, large $\coup$ causes slower mixing rates, which means that the performance of the sampler tends to peak significantly before the solution state has a higher probability than the zero cut state. This also serves as a good proxy for how close the sampler is to the model distribution, as the probability of reaching the ground state should peak when the MAX-CUT state has higher probability than the zero cut state if the sampled distribution is close to the model distribution.  In Figure \ref{fig:CoupVsZero} we show this phenomenon, where for $N_s = 70000$ we can see the regions of operation. For $N=50$ we can see the sampled distribution is very close to the model distribution, as the probability of reaching the ground state peaks when the ground state probability is larger than the zero cut probability and we are able to reach the ground state in almost all instances. For the $N=100$ and $N=150$ instances we can see the sampler is further away from the the model distribution as the probability of reaching the ground state decreases and the sampler performance peaks for smaller coupling values where $p_{soln} \ll p_{zero}$. 

\end{figure*}

\clearpage

\begin{figure*}

\begin{centering}
\includegraphics[width=\linewidth]{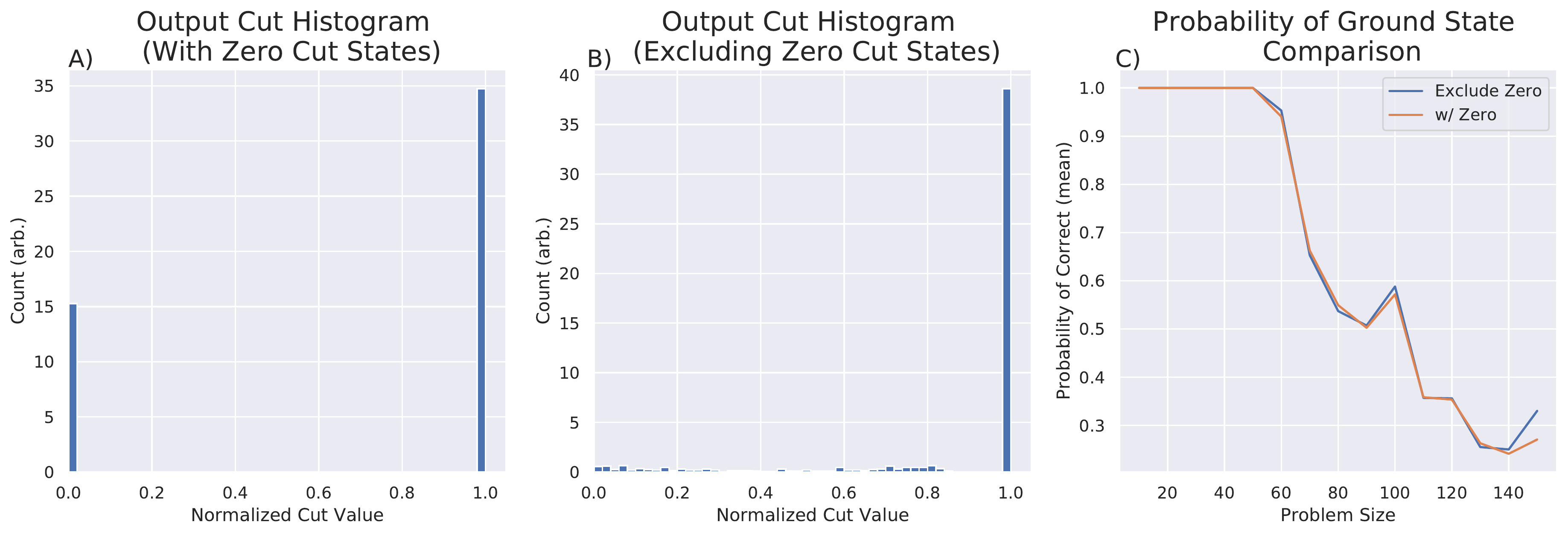}
\par\end{centering}
\caption{\label{fig:ExcludeZero}. \textbf{(Supplementary) Adding exclusion of zero cut states to hitting engine} \\ 
{\bf (A)\/} The output histogram of the stochastic sampler with hitting time engine for $N=150, N_s = 70000, \beta = 0.25, \coup=12$ shown after running 1000 times. We see that the output of the algorithm has clusters of solution at 0 cut and very close to optimal cut.  Referring to Supplementary Figure \ref{fig:CoupVsZero}, we can attribute this to the high probability assigned to the spurious zero cut state. 
{\bf (B)\/} We add an extra constraint in the hitting time engine to ignore all states that have an output cut of 0 (i.e. states with all 0 or all 1). The output is now concentrated very strongly around the optimal cut value, and all zero cut states are suppressed. 
{\bf (C)\/} Although the addition of the exclusion of zero states improves the average cut from the sampler, the probability of reaching the ground state remains mainly unchanged across all problem sizes. }

\vspace{2em}

Although the RBM has a relatively high probability of reaching the ground state ($\approx 30\%$ here), many of the sampling runs end by reaching a state with 0 cut value, where all nodes are on the same side of the cut and have the same value as shown above in Figure \ref{fig:ExcludeZero} A). This occurs when the coupling constraint is violated and the two physical nodes in the RBM are different for each logical node. Shown here for the optimal coupling value, we can see that the zero cut states still retain a very high probability causing a decrease in overall performance of the sampler. To combat this, we add an extra constraint into the hitting time engine to ignore all states of zero cut. This moves the average output cut up, improving the output of the sampler, shown in Figure \ref{fig:ExcludeZero} B), but doesn't effect the probability of reaching the ground state (shown in Figure \ref{fig:ExcludeZero} C)). 

\end{figure*}

\clearpage

\begin{figure*}
\begin{centering}
\includegraphics[width=\linewidth]{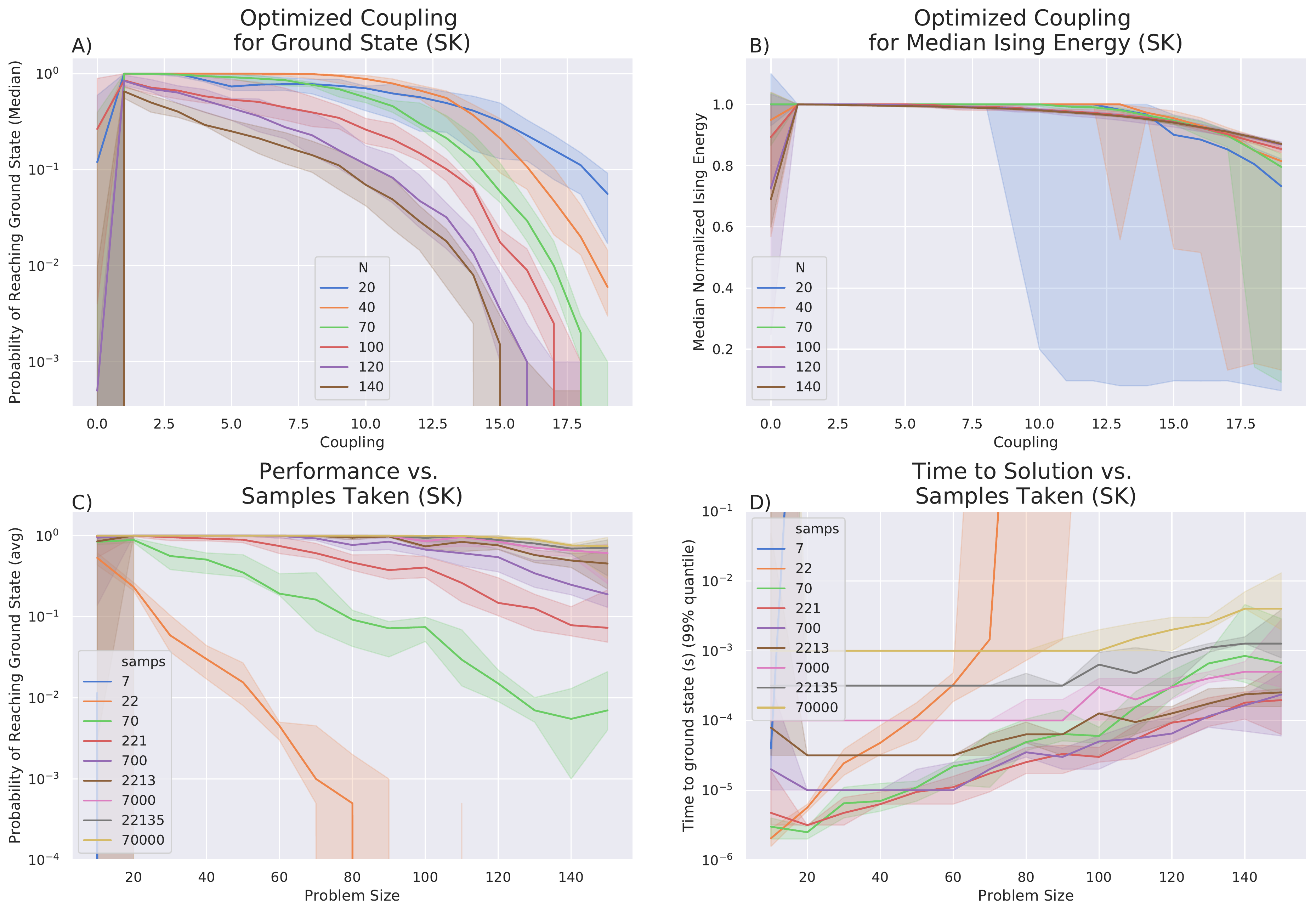}
\par\end{centering}
\caption{\label{fig:Param_SK}. \textbf{(Supplementary) Hyperparameter analysis on the Sherrington-Kirkpatrick (SK) Problem} \\ 
{\bf (A)\/} The coupling parameter for the SK problem is optimized at much lower values than the Max-Cut problem (closer to a coupling of 1 vs. a coupling of 12-13 on MAX-CUT). The SK problem has an inherent symmetry, as it has both +1 and -1 connections, causing the optimal coupling parameter to be lower for this type of problem. 
{\bf (B)\/} The optimal coupling value for maximizing the median cut follows the optimal coupling to find the ground state. Similar to the MaxCut problem, the median cut remains high and is not as sensitive to the coupling parameter as the probability of reaching the ground state. 
{\bf (C)\/} As above, in the MAX-CUT problem, the probability of reaching the ground state improves smoothly as more samples are taken. However, more samples takes more time, and we can optimize for the number of samples to take. The performance is significantly better on the SK problem, compared to the MAX-CUT problem when performed on the RBM. 
{\bf (D)\/} Using the time to solution framework outlined in the Results section above we can use the sampling performance in part (C) and find the time to solution for a given sample number. Here, we see fewer than 2000 samples should be taken across all problem sizes, much lower than the MAX-CUT problem. The minimum across all sample numbers is taken to find the global time to solution for the RBM. }
\end{figure*}

\clearpage
\begin{table*}

\centering
\begin{tabular}{ |p{2.5cm}||p{2.5cm}|p{2.5cm}|p{2.5cm}| p{2.5cm}| p{2.5cm}| } 
 \hline
 \thead{RBM Size \\ (Vis x Hid)} & \thead{LUT Usage \\ (Absolute)} & \thead{LUT Usage \\ (\%)} & \thead{FF Usage \\ (Absolute)} & \thead{FF Usage \\ (\%)} & \thead{Power \\ (W)} \\
 \hline\hline
 50x50 & 74389 & 6.29 & 40732 & 1.72  & 5.13\\  \hline
 100x100 & 267759 & 22.65  & 115188 & 4.87 & 5.35 \\  \hline
 150x150 & 575289 & 48.66 & 234680 & 9.93 & 5.60\\  \hline
 200x200 & 1007557 & 85.22 & 399183 & 16.88 & 5.86\\ \hline
 200x200 (no hitting engine) & 998407 & 84.45 & 393486 & 16.64 & 5.81 \\ \hline
\end{tabular}
\caption{\label{tab:utilization} (Supplementary) \textbf{FPGA Utilization Utilization numbers for FPGA and various RBM sizes}  
All usage numbers reported are for 9 bit weights and biases, including the hitting time engine (unless otherwise noted). The usage shows that the FPGA is not memory limited for the problem sizes we are interested in, but compute limited, as the LUT usage goes up much faster than the FF usage as the problem size grows. All weights and biases fit in on chip SRAM, allowing for fast access and data reuse. This also shows that the hardware overhead of the hitting time engine is minimal and should be included in designs to increase algorithmic performance.}
\end{table*}

\clearpage
\begin{figure*}
\begin{centering}
\includegraphics[width=\linewidth]{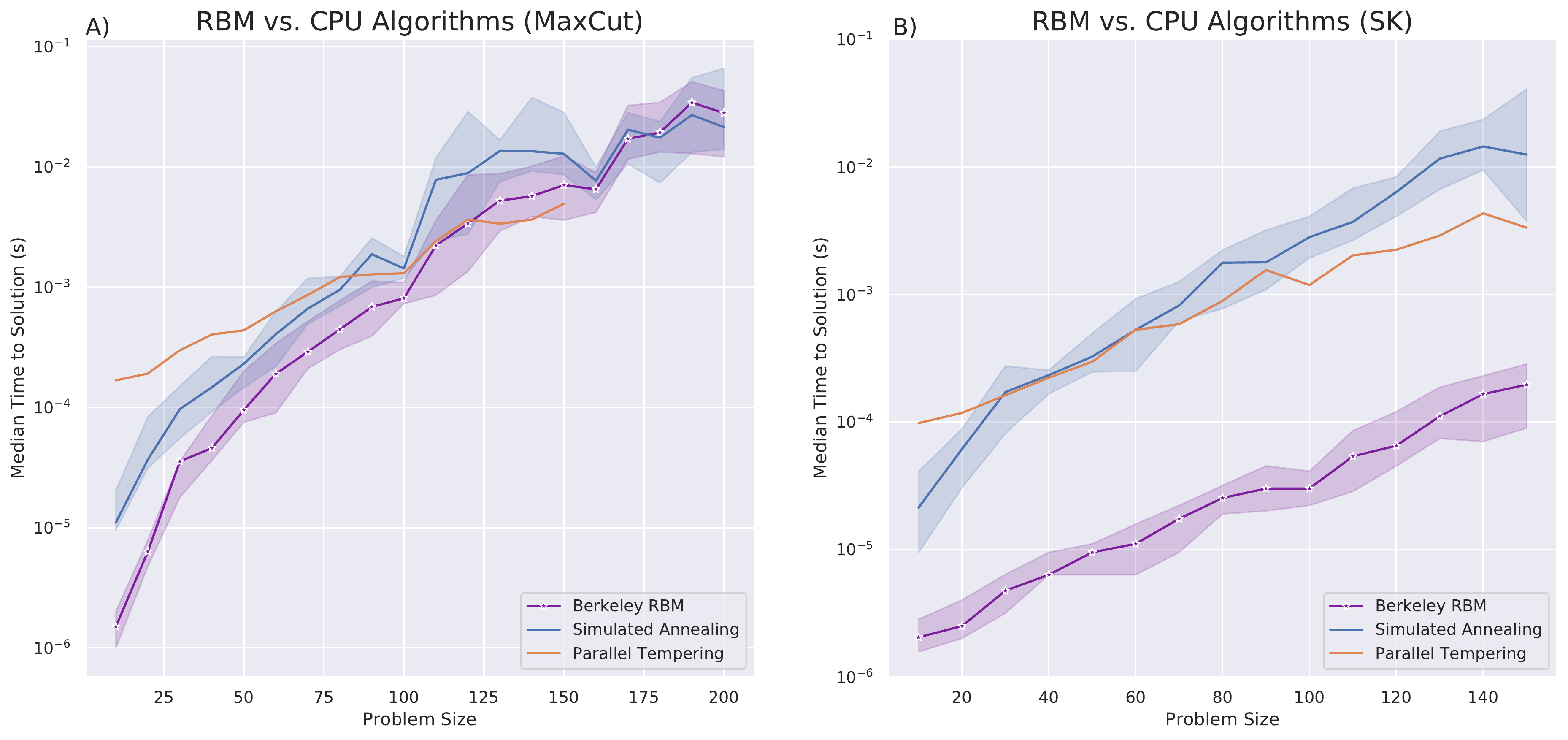}
\par\end{centering}
\caption{\label{fig:CPU}. \textbf{(Supplementary) Benchmarking against CPU algorithms} \\ 
{\bf (A)\/} The RBM performs competitively with two state of the art CPU algorithms for Ising Model problems, simulated annealing \cite{Isakov2014OptimizedGlasses, Kirkpatrick1983OptimizationAnnealing} and Parallel Tempering \cite{Hamerly2019ExperimentalAnnealer, Mandra2017Theagain, Mandra2018ADetection}. 
{\bf (B)\/} Comparison of the SK Problem against optimized CPU algorithms also yields constant factor speed improvement on the accelerated RBM. Across all problem instances we see a 10-20x speed improvement due to the hardware acceleration. 
}

\vspace{2em}
Simulated Annealing is performed on a Xeon E5-2620 processor using the code from \cite{Isakov2014OptimizedGlasses}, while the Parallel Tempering results are copied from \cite{Hamerly2019ExperimentalAnnealer} using the NASA/TAMU Unified Framework for Optimization, running on a Xeon E5-1650 v2 processor.  The RBM performs better than both algorithms for small problem instances on the MAX-CUT problem (closer to 5x improvement), but converges for the larger instances presented in the dataset. Although parallel tempering narrowly outperforms the RBM on large MAX-CUT instances, the empirically seen $\mathcal{O}(e^{\sqrt{N}})$ scaling of the RBM is asymptotically favorable to the $\mathcal{O}(e^{N})$ scaling of the parallel tempering algorithm. The RBM is able to outperform both these algorithms on the SK problem across all problem instances, demonstrating a performance advantage for problems with full connectivity. The RBM performs competitively with these state of the art algorithms, but more work needs to be done to increase the performance relative to these baselines. Many of the optimizations used in the simulated annealing (pre-computing energies, using weight sparsity, efficient random number generation) can be implemented with the RBM as well to increase its performance. Parallel tempering can also be added to the RBM sampling algorithm to yield improved sampling, which we expect to improve the overall algorithm scaling and performance. \cite{Desjardins2010ParallelMachines, Cho2010ParallelMachines}
 
\end{figure*}


\end{document}